\documentclass{midl} % Include author names
 
\usepackage{booktabs}
\usepackage{amsfonts}
\usepackage{bm}
 
\jmlrproceedings{MIDL}{Medical Imaging with Deep Learning}
\jmlrpages{}
\jmlryear{2026}
 
\jmlrworkshop{Short Paper Track}
\jmlrvolume{}
\editors{}

\title[CASA]{Physics-Grounded Adversarial Stain Augmentation\\with Calibrated Coverage Guarantees}

\midlauthor{\Name{Mingi Hong} \orcid{0009-0009-4121-5765} \Email{mghong@mtsco.co.kr}\\
\addr 434, Samseong-ro, Gangnam-gu, Seoul, Republic of Korea
}

\begin{document}

\maketitle

\begin{abstract}
Stain variation across hospitals degrades histopathology models at deployment.
Existing augmentation methods perturb color spaces with arbitrary hyperparameters, lacking both a principled budget and coverage guarantees for unseen centers.
We propose \textbf{C}alibrated \textbf{A}dversarial \textbf{S}tain \textbf{A}ugmentation (\textbf{CASA}), which performs adversarial augmentation in the Macenko stain parameter space with a budget calibrated from multi-center statistics via the DKW inequality.
On Camelyon17-WILDS (5 seeds), CASA achieves $93.9\% \pm 1.6\%$ slide-level accuracy---outperforming HED-strong ($88.4\% \pm 7.3\%$), RandStainNA ($85.2\% \pm 6.7\%$), and ERM ($63.9\% \pm 11.3\%$)---with the highest worst-group accuracy ($84.9\% \pm 0.9\%$) among all 10 compared methods.
\end{abstract}

\begin{keywords}
Histopathology, Domain Generalization, Adversarial Augmentation, Distributionally Robust Optimization
\end{keywords}

\section{Introduction}
 
Hematoxylin and Eosin (H\&E) staining varies across hospitals due to differences in reagent concentrations, protocols, and scanners.
This variation causes models trained at one center to degrade at another, limiting clinical deployment \cite{tellez2019quantifying}.
 
Existing approaches fall into two categories.
\textit{Stain normalization} transforms images to a reference template but introduces artifacts and depends on template selection \cite{macenko2009method}.
\textit{Stain augmentation} increases color diversity during training; Tellez et al.\ showed that random perturbation in HED space improves generalization \cite{tellez2019quantifying}, and RandStainNA bridges augmentation with normalization in LAB space \cite{shen2022randstainna}.
However, random methods require manual selection of perturbation strength ($\sigma$), and this choice affects OOD performance: we observe an 8 percentage point gap between HED-light ($\sigma{=}0.05$, $80.4\%$) and HED-strong ($\sigma{=}0.2$, $88.4\%$) on Camelyon17-WILDS, with no principled criterion for selecting $\sigma$.
 
Recent adversarial augmentation methods---AdvST \cite{zheng2024advst} and Zhong et al.\ \cite{zhong2022adversarial}---learn worst-case perturbations via PGD but operate in generic color spaces (HSV, contrast) with arbitrary budgets unconnected to the physical staining process.
 
If instead the model trains on the \emph{hardest} stain variation within a realistic range, it should generalize to any center whose stain falls within that range.
We propose \textbf{CASA}, which realizes this idea: (1) adversarial augmentation in the \textit{physics-grounded} Macenko stain parameter space, where perturbations correspond to realistic stain variations; and (2) an adversarial budget \textit{calibrated} from multi-center data with a statistical coverage guarantee.

\section{Method}
 
\paragraph{Stain decomposition.}
The Beer-Lambert law models H\&E image formation: for each pixel $j$, $\mathbf{I}_j = I_0 \exp(-\mathbf{W}\mathbf{h}_j)$, where $\mathbf{W} \in \mathbb{R}^{3\times 2}$ contains unit-norm stain basis vectors (Hematoxylin and Eosin columns) and $\mathbf{h}_j \in \mathbb{R}^{2}$ contains the corresponding stain concentrations.
We extract $\mathbf{W}$ via the Macenko method \cite{macenko2009method}, building on the color deconvolution framework of Ruifrok and Johnston \cite{ruifrok2001quantification}, and obtain $\mathbf{h}$ by least-squares inversion.
 
\paragraph{Training loop.}
Each iteration solves a min-max problem.
The \textit{inner maximization} finds the worst-case stain perturbation $(\bm{\delta}_W^*, \bm{\delta}_h^*)$ via $K$-step PGD:
\begin{equation}\label{eq:inner}
\max_{\|\bm{\delta}_W\| \leq \tau_W,\; \|\bm{\delta}_h\| \leq \tau_H} \; \mathcal{L}\!\Big(f_\theta\!\big(I_0 e^{-(\mathbf{W}_{\text{ref}} + \bm{\delta}_W)\,(\mathbf{h}_0 \odot (1 + \bm{\delta}_h))}\big),\; y\Big)
\end{equation}
where $\bm{\delta}_W$ perturbs stain vector directions (projected onto a spherical cap to preserve unit norm) and $\bm{\delta}_h$ scales concentrations element-wise.
Since $\tau_H < 1$, the factor $(1 + \delta_{h,k})$ remains strictly positive for the final perturbation; intermediate PGD iterates are clamped to $\delta_{h,k} \geq -1$ to maintain non-negativity throughout.
The reference stain matrix $\mathbf{W}_{\text{ref}}$ and initial concentrations $\mathbf{h}_0$ are precomputed once per batch from the Macenko decomposition.
After $K$ steps, the augmented image is reconstructed via the Beer-Lambert inverse and passed to the \textit{outer minimization}, which updates $\theta$ by SGD on $\mathcal{L}(f_\theta(\tilde{\mathbf{I}}), y)$.
The model trains on the worst-case stain perturbation within the calibrated budget at every iteration.
 
\paragraph{Calibrated budget.}
The budgets $\tau_W$ and $\tau_H$ are not hyperparameters---they are estimated from training data.
Given $n$ images from the training centers:
(i) extract per-image stain matrices $\mathbf{W}_i$ and concentrations $\mathbf{h}_i$;
(ii) compute the angular deviation $\alpha_i = \angle(\mathbf{W}_i, \bar{\mathbf{W}})$ and concentration ratio $r_i = q_{99}(\mathbf{h}_i) \,/\, \overline{q_{99}(\mathbf{h})}$, where $q_{99}$ denotes the 99th percentile;
(iii) set $\tau_W$ and $\tau_H$ as the $(1{-}\delta{+}\varepsilon_n)$-empirical quantiles of $\{\alpha_i\}$ and $\{|r_i - 1|\}$ respectively, where $\varepsilon_n$ is the DKW correction below.
On Camelyon17-WILDS with $n{=}1{,}000$ training images, this yields $\tau_W{=}0.353$ rad ($20.2^\circ$) and $\tau_H{=}0.987$.
 
\paragraph{Coverage guarantee.}
The DKW inequality \cite{dvoretzky1956asymptotic,massart1990tight} bounds the deviation of the empirical CDF $F_n$ from the true CDF $F$:
\begin{equation}\label{eq:coverage}
\Pr\!\left[\sup_t |F_n(t) - F(t)| > \varepsilon_n\right] \leq 2e^{-2n\varepsilon_n^2}
\end{equation}
Since we apply the bound to both $\tau_W$ and $\tau_H$, a union bound requires each to hold at level $\beta/2$, giving $\varepsilon_n = \sqrt{\ln(4/\beta)/(2n)}$.
We take the $(1{-}\delta{+}\varepsilon_n)$-empirical quantile as each budget, which ensures that the true $(1{-}\delta)$-quantile is covered with probability $\geq 1{-}\beta$.
With $\delta{=}\beta{=}0.05$ and $n{=}1{,}000$: $\varepsilon_n {=} 0.047$, giving a quantile level of $0.997$.
Each stain parameter of a new center thus falls within its budget with $\geq 95\%$ confidence.

\section{Experiments}
 
\paragraph{Setup.}
We evaluate on Camelyon17-WILDS \cite{koh2021wilds}: binary tumor classification across 5 hospitals (train: 0,3,4; val: 1; test: 2; 302K patches).
All methods use DenseNet-121 trained from scratch, SGD (lr$=$0.001, wd$=$0.01), batch size 32, 10 epochs, 5 seeds.
CASA uses $K{=}5$ PGD steps.
We report the WILDS official slide-level macro accuracy (\texttt{acc\_avg}).
Baselines include stain augmentation (HED-strong/light \cite{tellez2019quantifying}, RandStainNA \cite{shen2022randstainna}, Macenko-norm \cite{macenko2009method}), adversarial methods (AdvST \cite{zheng2024advst}, Zhong et al.\ \cite{zhong2022adversarial}), domain adaptation (DANN \cite{ganin2016domain}), and bilevel stain optimization in the same parameter space as CASA.
 
\paragraph{Results.}
 
\begin{table}[htbp]
\floatconts
  {tab:main}
  {\caption{Camelyon17-WILDS test results (DenseNet-121, 5 seeds).}}
  {\begin{tabular}{lcc}
  \toprule
  \bfseries Method & \bfseries acc\_avg & \bfseries acc\_wg \\
  \midrule
  \textbf{CASA (ours)} & $\mathbf{93.9 \pm 1.6}$ & $\mathbf{84.9 \pm 0.9}$ \\
  HED-strong ($\sigma{=}0.2$) & $88.4 \pm 7.3$ & $76.1 \pm 13.5$ \\
  RandStainNA            & $85.2 \pm 6.7$ & $57.6 \pm 21.3$ \\
  HED-light ($\sigma{=}0.05$) & $80.4 \pm 6.6$ & $66.1 \pm 10.3$ \\
  Macenko-norm           & $78.0 \pm 14.7$ & $46.0 \pm 17.5$ \\
  Bilevel-Stain          & $71.4 \pm 12.7$ & $43.2 \pm 14.5$ \\
  DANN                   & $69.8 \pm 4.1$ & $53.1 \pm 5.8$ \\
  Zhong et al.           & $64.8 \pm 9.8$ & $46.8 \pm 9.3$ \\
  ERM                    & $63.9 \pm 11.3$ & $36.1 \pm 9.4$ \\
  AdvST                  & $56.4 \pm 7.1$ & $31.7 \pm 14.9$ \\
  \bottomrule
  \end{tabular}}
\end{table}
 
CASA ranks first in both accuracy ($93.9\%$) and worst-group stability ($\pm 0.9\%$).
Three findings emerge from \tableref{tab:main}:
 
\textit{(1) Stain-specific $>$ generic adversarial.}
AdvST ($56.4\%$) perturbs HSV, contrast, and sharpness---it scores below ERM ($63.9\%$).
Zhong et al.\ ($64.8\%$) operates on feature-level style and only matches ERM.
Adversarial augmentation requires stain-specific parameterization to help rather than hurt.
 
\textit{(2) Adversarial $>$ random in the same space.}
CASA ($93.9\%$) outperforms HED-strong ($88.4\%$) by $5.5$ points with $4{\times}$ lower \texttt{acc\_avg} variance.
Bilevel optimization ($71.4\%$), which minimizes rather than maximizes the inner objective in the same parameter space, scores 22.5 points lower---worst-case perturbations generalize better than ``helpful'' ones.
 
\textit{(3) Calibrated budget removes hyperparameter sensitivity.}
HED-strong ($\sigma{=}0.2$) and HED-light ($\sigma{=}0.05$) differ by 8 points, yet no criterion exists for choosing $\sigma$.
CASA derives its budget from the data; no manual tuning is needed.

\section{Conclusion}
 
Two design choices make CASA effective where generic adversarial methods fail: grounding perturbations in the physical stain parameter space, and calibrating the budget from multi-center data instead of tuning it by hand.
Because the only domain-specific components are the stain decomposition and the calibration set, the same min-max formulation transfers to any staining protocol whose forward model is known.

\bibliography{references}

\end{document}